\documentclass[english,aps,prl,twocolumn,showpacs,reprint]{revtex4}
\usepackage[T1]{fontenc}
\usepackage[latin9]{inputenc}
\usepackage{amsbsy}
\usepackage{amstext}
\usepackage{amssymb}
\usepackage{graphicx}
\usepackage{esint}

\makeatletter
\@ifundefined{textcolor}{}
{%
 \definecolor{BLACK}{gray}{0}
 \definecolor{WHITE}{gray}{1}
 \definecolor{RED}{rgb}{1,0,0}
 \definecolor{GREEN}{rgb}{0,1,0}
 \definecolor{BLUE}{rgb}{0,0,1}
 \definecolor{CYAN}{cmyk}{1,0,0,0}
 \definecolor{MAGENTA}{cmyk}{0,1,0,0}
 \definecolor{YELLOW}{cmyk}{0,0,1,0}
}

\makeatother

\usepackage{babel}
\begin{document}

\title{Enhancement and reduction of one-dimensional heat conduction with
correlated mass disorder }

\author{Zhun-Yong Ong}

\email{ongzy@ihpc.a-star.edu.sg}

\affiliation{Institute of High Performance Computing, A{*}STAR, Singapore 138632,
Singapore}

\author{Gang Zhang}

\email{zhangg@ihpc.a-star.edu.sg}

\affiliation{Institute of High Performance Computing, A{*}STAR, Singapore 138632,
Singapore}
\begin{abstract}
Short-range order in strongly disordered structures plays an important
role in their heat conduction property. Using numerical and analytical
methods, we show that short-range spatial correlation (with a correlation
length of $\Lambda_{m}$) in the mass distribution of the one-dimensional
(1D) alloy-like random binary lattice leads to a dramatic enhancement
of the high-frequency phonon transmittance but also increases the
low-frequency phonon opacity. High-frequency semi-extended states
are formed while low-frequency modes become more localized. This results
in ballistic heat conduction at finite lengths but also paradoxically
higher thermal resistance that scale as $\sqrt{\Lambda_{m}}$ in the
$L\rightarrow\infty$ limit. We identify an emergent crossover length
($L_{c}$) below which the onset of thermal transparency appears.
The crossover length is linearly dependent on but is two orders of
magnitude larger than $\Lambda_{m}$. Our results suggest that the
phonon transmittance spectrum and heat conduction in a disordered
1D lattice can be controlled via statistical clustering of the constituent
component atoms into domains. They also imply that the detection of
ballistic heat conduction in disordered 1D structures may be a signature
of the intrinsic mass correlation at a much smaller length scale.
\end{abstract}


\maketitle

\section{Introduction}

Phonon-mediated heat conduction in low-dimensional nanostructures
is a transport phenomenon of fundamental and applied interest.\ \cite{MMaldovan:PRL13_Narrow}
In particular, the manipulation of the thermal conductivity in nanomaterials
may enable the realization of potential applications in thermoelectric
devices, solid state refrigeration and thermal cloaking.\ \cite{THan13:SciRep13}
One approach is to use a low-dimensional material such as nanowires\ \cite{JZou:JAP01,DLi:APL03_SiNW,AHochbaum:Nature08_Thermoelectric,DLNika:APL13_SiGeNW}
in which confinement alters the intrinsic phononic properties (e.g.\emph{,}
anisotropy, dispersion and mean free path). The other is to use a
bulk material such as Si or Ge and modify its thermal conductivity
($\kappa$) via nanostructuring. This includes the creation of periodic
patterns (e.g.\emph{,} Si-Ge superlattices\ \cite{MVSimkin:PRL00,DLi:APL03_SiGeNW,CDames:JAP04_SiGe,PChen:PRL14}
and nanopore arrays\ \cite{PEHopkins:NL10_Reduction,JYu:NatNano10_Nanomesh,AJain:PRB13}),
the use of resonant superstructures\ \cite{BLDavis:PRL14}, and alloying.\ \cite{GHZhu:PRL09}
The latter appears to be one of the more promising options for obtaining
the reduced thermal conductivity needed for efficient and cost-effective
thermoelectric devices because an abrupt decrease in $\kappa$ can
be observed upon the introduction of a small alloy concentration.\ \cite{GHZhu:PRL09}
Alloying can also be combined with confinement to further reduce the
thermal conductivity.

The low thermal conductivity in alloys can be attributed in part to
mode localization which stems from the random placement of the component
atoms and impedes phonon propagation. For the finite three-dimensional
disordered harmonic solid, it has been shown that only delocalized
modes contribute significantly to the heat current.\ \cite{PBAllen:PRB93_Thermal,JLFeldman:PRB93_Thermal}
In one dimension (1D) in particular, disorder results in the localization
of all the modes in in the thermodynamic ($L\rightarrow\infty$) limit\ \cite{HMatsuda:PTP70_Localization,KIshii:PTPS73,ADhar:PRL01_DisorderedHarmonicChain}
although for a finite system, a fraction of the states would always
be sufficiently extended to contribute to the heat current.\ \cite{ADhar:AdvPhys08_Heat}
Valuable insights into the interplay between localization and propagation
in 1D can be gleaned within the framework of the disordered harmonic
chain (DHC), the simplest model of a disordered 1D structure that
has been frequently used to understand the effect of disorder on heat
conduction.\ \cite{RJRubin:JMP68_Transmission,ADhar:PRL01_DisorderedHarmonicChain,ZYOng:Manuscript14_MassDisorder}
Broadly speaking, we know that disorder leads to the exponential attenuation
of the phonon transmittance $\Xi$, i.e., $\Xi(\omega,L)\sim\exp[-L/\lambda(\omega)]$,
where $\omega$ and $\lambda$ are the frequency and attenuation length,
respectively, in analogy to the Beer-Lambert law in optics,\ \cite{ZYOng:Manuscript14_MassDisorder}
and is a direct consequence of localization.\ \cite{HMatsuda:PTP70_Localization,KIshii:PTPS73,ADhar:PRL01_DisorderedHarmonicChain}
It follows from the known $\lambda\propto\omega^{-2}$ relationship\ \cite{ZYOng:Manuscript14_MassDisorder}
that the thermal conductivity scales as $\lim_{L\rightarrow\infty}\kappa\propto L^{0.5}$
and $\lim_{L\rightarrow\infty}\kappa\propto\langle\delta m\rangle^{-1}$,
where $L$ and $\langle\delta m\rangle$ are respectively the length
and mass fluctuation, for free boundary conditions.\ \cite{HMatsuda:PTP70_Localization,WMVisscher:PTP71,RJRubin:JMP71_Abnormal,WLGreer:JMP72_Quantum,SLepri:PhysRep03}. 

On the other hand, these semirigorous results have only been established
for the case of heat conduction with purely random (uncorrelated)
mass disorder, in which the position of the mass fluctuation is uncorrelated
across different atomic sites.\ \cite{RJRubin:JMP68_Transmission,HMatsuda:PTP70_Localization,KIshii:PTPS73,ADhar:PRL01_DisorderedHarmonicChain}
In an alloy, this disorder manifests itself as the uncorrelated placement
of the constituent atoms. However, when there is statistical clustering
of the atoms in the form of domains, the position of the atoms and
the spatial distribution of mass become correlated, introducing short-range
order and modifying the localization phenomenon. A considerable amount
of work has been done by de Moura and co-workers\ \cite{FABFDeMoura:PRB03_DelocalizationCorrelatedRandomMasses,FABFDeMoura:PRB06_Aperiodic}
who showed by using numerical simulations that \emph{long-range} correlation
leads to the formation of low-energy extended states in the DHC. Duda
and co-workers have also studied the effects of atomic ordering on
the thermal conductivity by using molecular dynamics simulations.\ \cite{JCDuda:JPCM11,JCDuda:JHT12_Controlling}
Using an analytical approach, Herrera-Gonzalez and co-workers\ \cite{IFHerreraGonzalez:EPL10_Anomalous}
also studied the relationship between the scaling of the thermal conductivity
with system size and \emph{long-range} correlated isotopic disorder.
They demonstrate that specific long-range correlations can suppress
or enhance the heat current contribution of vibrational modes in pre-defined
frequency windows. However, their results are confined to the case
of weak isotopic disorder and weak coupling between the lattice and
the heat baths, owing to the nature of their perturbative approximations.
Their numerical simulations are also limited to systems with $N\leq10^{3}$,
where $N$ is the number of atoms. Here, we adopt a numerical approach
in this paper to more fully and systematically explore the implications
of \emph{short-range} \emph{correlated mass disorder} for heat conduction,
especially with regard to its dependence on length ($N\leq1.6\times10^{5}$)
and the extent of the short-range order in mass correlation.

In this paper, we report the effect of \emph{short-range} mass correlation
(or correlated disorder, for short) on phonon transmittance and heat
conduction through a random binary lattice (RBL), an alloy-like realization
of the DHC. We restrict ourselves to the coherent processes and neglect
anharmonic effects from inelastic scattering between phonons. This
allows us to isolate the effects of disorder on heat conduction in
micrometer-scale systems, and we expect our results to be applicable
to 1D alloy nanostructures for which the reduced thermal conductivity
can be traced to mass disorder.\ \cite{JMLarkin:JAP13_Thermal,THori:JAP13_Phonon}
In the following sections, we first describe the 1D lattice model
and how correlated disorder is generated for the lattice. The short-range
order in the mass correlation, in the form of an exponentially decaying
spatial correlation, is characterized by a correlation length of $\Lambda_{m}$.
Next, the phonon transmittance is calculated for the uncorrelated
and correlated model. We find from numerical simulations that the
short-range correlation in the mass distribution, through the formation
of domains, leads to an emergent crossover length scale ($L_{c}$)
below which the system is effectively transparent over a wide phonon
frequency range and conducts heat ballistically. At very low frequencies,
mass correlation decreases the attenuation length while at higher
frequencies, the attenuation length actually increases. Thus, when
$L$ exceeds $L_{c}$, one sees a rapid onset of phononic opacity,
resulting in a dramatic increase in the thermal resistance exceeding
that in the DHC with uncorrelated mass disorder. More intriguingly,
$L_{c}$ is two orders of magnitude larger than $\Lambda_{m}$. We
find a formula that connects the mass correlation length and the increase
in thermal resistance in long chains. Finally, we discuss the connection
between our purely 1D results to heat conduction in more realistic
systems, and how correlated mass disorder can affect heat conduction
at length scales much larger than the correlation length. We also
interpret the recent findings of ballistic heat conduction in SiGe-alloy
nanowires in Ref.\ \cite{TKHsiao:NatNano13_Observation} in light
of our results.

\section{Methodology}

\subsection{Random binary lattice model}

We choose as our model system the familiar disordered harmonic chain
first studied by Dyson,\ \cite{FJDyson:PR53_DisorderedLinearChain}
which we also used in our earlier paper.\ \cite{ZYOng:Manuscript14_MassDisorder}
Our system differs from that used in other papers in that the atomic
mass is not treated as a continuous random variable.\ \cite{FABFDeMoura:PRB03_DelocalizationCorrelatedRandomMasses,JDBodyfelt:PRE13_ScalingTheory}
Rather, we have a random binary lattice, where the positions of the
component atoms are set probabilistically so that the lattice resembles
more realistic \emph{alloy} systems with mass disorder. For simplicity,
only adjacent atoms are coupled. The atoms are only permitted to move
longitudinally and no attempt is made to include any anharmonic interaction
in our model. 

Our system consists of three parts. In the middle, there is a finite-size
``conductor'' of $N$ equally spaced atoms. On either side, there
is a homogeneous lead, i.e., a semi-infinite chain with no mass disorder.
In effect, we have a finite disordered system embedded in an \emph{infinite}
homogeneous 1D lattice, which also corresponds to a finite disordered
harmonic chain with free boundaries.\ \cite{RJRubin:JMP71_Abnormal,ADhar:PRL01_DisorderedHarmonicChain}
The homogeneous leads act as heat reservoirs and are coupled to the
conductor via the same harmonic spring terms as those between the
atoms in the disordered lattice. Coupling between adjacent atoms is
governed by the harmonic spring term $V(x_{i},x_{i+1})=\frac{1}{2}k(x_{i}-x_{i+1})^{2}$
where $k$ is the spring constant and $x_{i}$ is the displacement
of the $i$-th atom from its equilibrium position. In the RBL, there
are two species of atoms which we label \emph{A} and \emph{B}. Species
\emph{A} is taken to be the substitutional impurity and exists only
within the conductor. The rest of the atoms in the conductor and the
leads are of species \emph{B}. We set the mass of the \emph{A} (\emph{B})
atoms to $m_{A}=9.2\times10^{-26}$ kg ($m_{B}=4.6\times10^{-26}$
kg, the mass of a Si atom), the spring constant to $k=32$ Nm$^{-1}$
(the approximate strength of the Si-Si bond) and the interatomic spacing
to $a=0.55$ nm. Our choice of parameters follows that in Ref.\ \cite{WZhang:NHT07}.
The maximum simulated chain length is $N=1.6\times10^{5}$ or $L=Na=88$
$\mu$m.

\subsection{Short-range order in lattice}

\begin{figure}
\includegraphics[width=8.5cm]{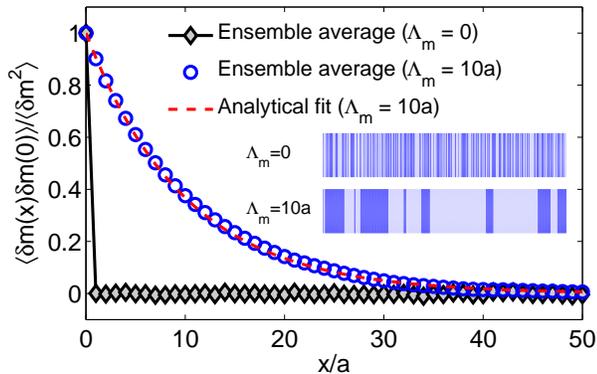}

\caption{(Color online) Plot of the ensemble-average mass correlation function
for uncorrelated (diamond symbol, $\Lambda_{m}=0$) and correlated
(circle symbol, $\Lambda_{m}=10a$) mass disorder at $c_{A}=c_{B}=0.5$.
The mass correlation is calculated numerically from 100 realizations
of disorder. The analytical fit based on Eq.\ (\ref{Eq:ExponentialMassCorrelation})
is also shown (dashed line). A representative schematic of their mass
distributions is shown in the inset. The correlated distribution consists
of much larger domains of \emph{A} and \emph{B} atoms.}

\label{Fig:MassDistribution}
\end{figure}

We define the mass fluctuation correlation function as $\left\langle \delta m(x)\delta m(x')\right\rangle =\left\langle m(x)m(x')\right\rangle -\left\langle m(x)\right\rangle ^{2}$,
where $m(x)$ is the atomic mass at site $x$. It measures the short-range
correlation in mass fluctuation (or mass correlation for short). In
RBL, we set the distribution of the constituent atoms such that the
mass correlation has the exponential form 
\begin{equation}
\left\langle \delta m(x)\delta m(x')\right\rangle =c_{A}c_{B}\Delta m^{2}\exp(-|x-x'|/\Lambda_{m})\ ,\label{Eq:ExponentialMassCorrelation}
\end{equation}
where $\Delta m=|m_{A}-m_{B}|$ and $c_{A}$ ($c_{B}$) is the concentration
of \emph{A} (\emph{B}) atoms. For pure random disorder ($\Lambda_{m}=0$),
the mass correlation function is $\left\langle \delta m(x)\delta m(0)\right\rangle =c_{A}c_{B}(m_{A}-m_{B})^{2}\delta_{x,0}$.
To generate the lattice configuration with the mass correlation in
Eq.\ (\ref{Eq:ExponentialMassCorrelation}), we first divide the
conductor into smaller domains alternating between having all \emph{A}
or all \emph{B} atoms. The number of sites in each domain of \emph{A}
atoms ($d_{A}$) is generated according to the probability distribution
$P(d)=\langle d_{A}\rangle^{-1}\exp(-d/\langle d_{A}\rangle)$ where
$\langle d_{A}\rangle=\Lambda_{m}[a(1-c_{A})]^{-1}$ is the average
type-\emph{A} domain size. The size of each type-\emph{B} domain ($d_{B}$)
is similarly defined. For simplicity's sake, we set $c_{A}=c_{B}=0.5$.
In the uncorrelated disorder case, we do not create smaller domains
but instead set the probability of each site having an \emph{A} (\emph{B})
atom to $c_{A}$ ($c_{B}$). Figure\ \ref{Fig:MassDistribution}
shows the normalized mass correlation function $\left\langle \delta m(x)\delta m(x')\right\rangle /\left\langle \delta m(x)^{2}\right\rangle $
for uncorrelated and correlated ($\Lambda_{m}=10a$) mass disorder.
A schematic representation of the mass distribution can also be seen
in the inset of Fig.\ \ref{Fig:MassDistribution}.

\subsection{Phonon transmittance calculation}

There are two approaches to computing the phonon transmittance through
the RBL. The first is the well-known nonequilibrium Green's function
method\ \cite{WZhang:NHT07,NMingo:Springer09} in which the transmittance
is given by $\Xi(\omega)=\text{Tr}(\boldsymbol{\Gamma_{L}G\Gamma_{R}G^{\dagger}})$,
where $G$ is the nonequilibrium Green's function and $\Gamma_{L}$
($\Gamma_{R}$) is the term coupling the left (right) lead to the
conductor. More details of this method can be found in Refs.\ \cite{WZhang:NHT07,ZYOng:Manuscript14_MassDisorder}.
The second approach is via the eigenvalue of the products of transfer
matrices,\ \cite{FABFDeMoura:PRB03_DelocalizationCorrelatedRandomMasses,FABFDeMoura:PRB06_Aperiodic}
which we describe in Appendix\ \ref{Appendix:AnalyticalEstimate}.
In either case, the ensemble average of the transmittance $\langle\Xi(\omega)\rangle$
is taken over 100 independent realizations of mass disorder.

\begin{figure}
\includegraphics[width=7.5cm]{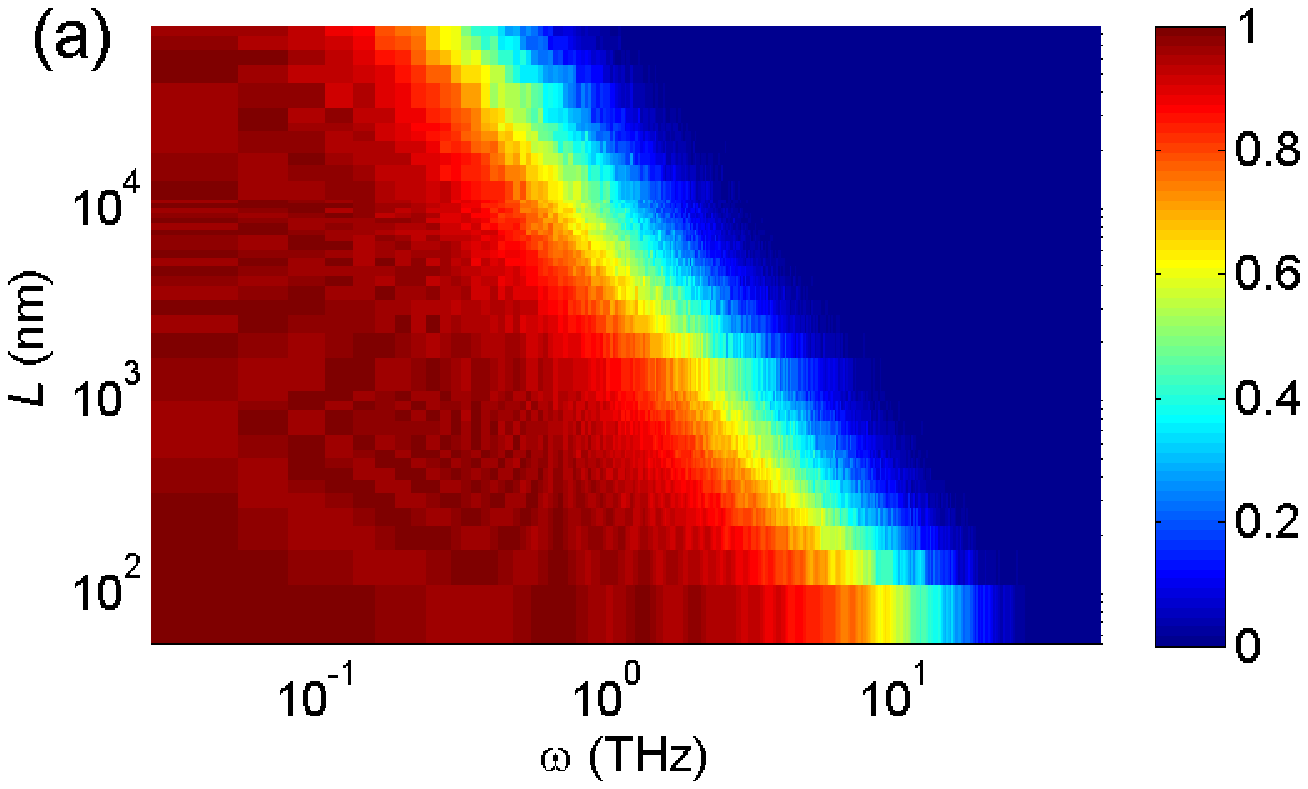}

\includegraphics[width=7.5cm]{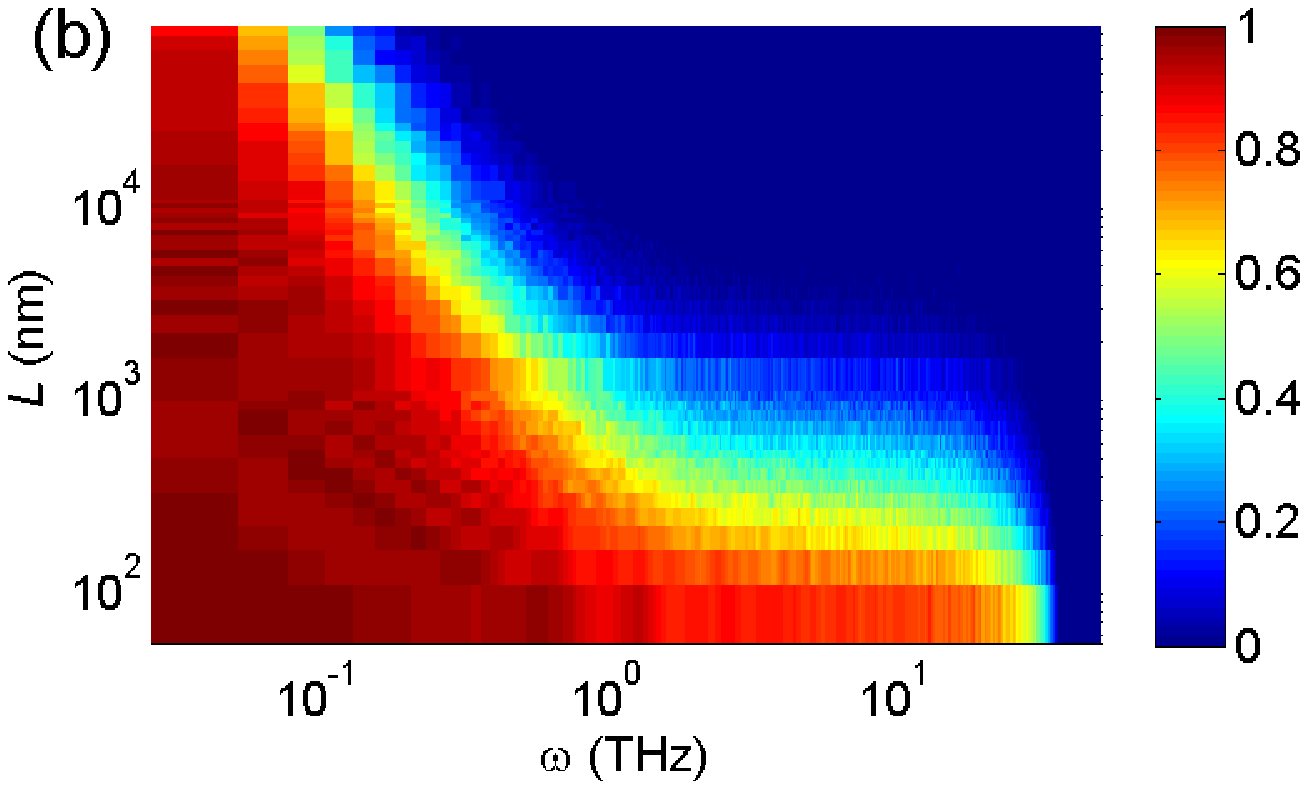}

\includegraphics[width=7.5cm]{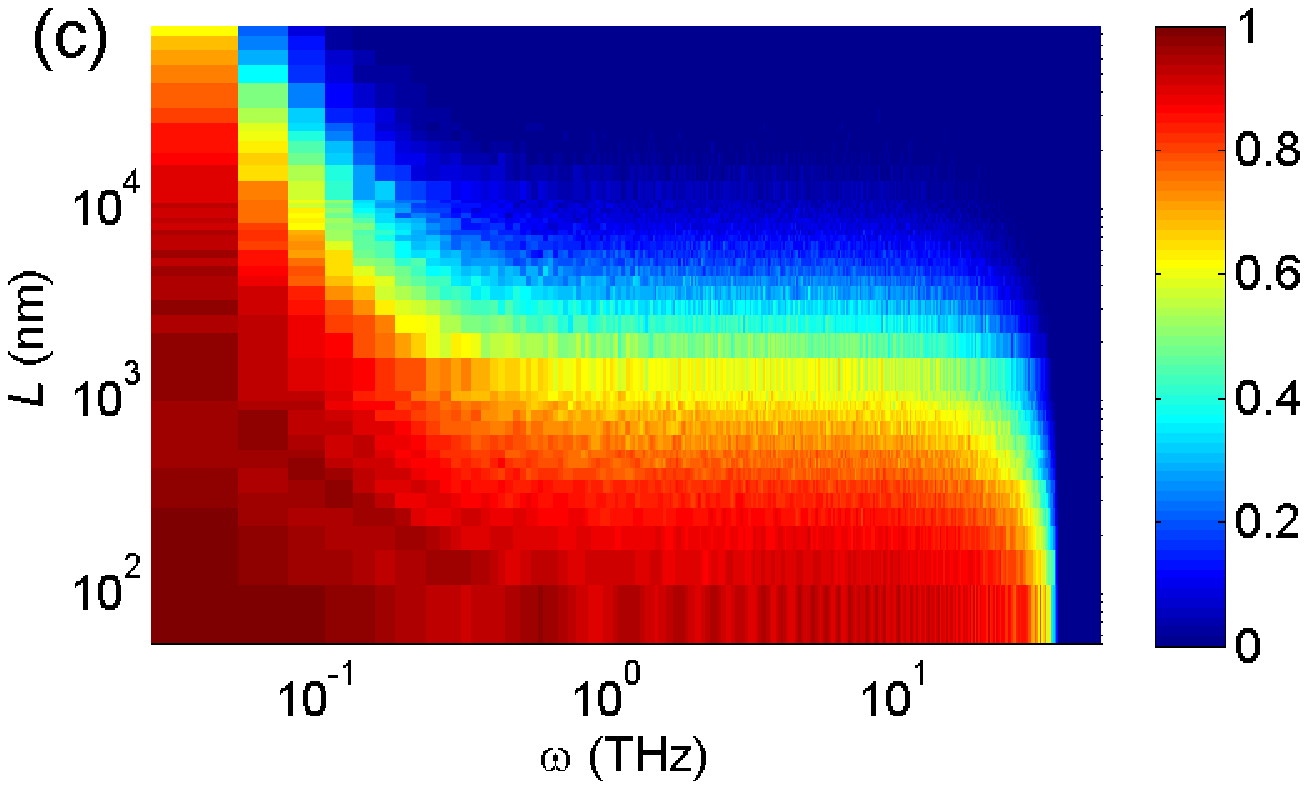}

\caption{(Color online) Plot of transmittance function $\langle\Xi(\omega,L)\rangle$
as a function of chain length $L$ and frequency $\omega$ for (a)
$\Lambda_{m}=0$, (b) $\Lambda_{m}=10a$, and (c) $\Lambda_{m}=50a$.
In the uncorrelated case ($\Lambda_{m}=0$), the transmittance scales
as $\langle\Xi(\omega,L)\rangle\sim\exp[-L/\lambda(\omega)]$ where
$\lambda\propto\omega^{-2}$. However, for the correlated mass distribution
in (b) and (c), the transmittance has the form $\langle\Xi(\omega,L)\rangle\sim\exp[-L/\lambda(\omega)]$
but the attenuation length no longer scales as $\lambda\propto\omega^{-2}$.
In the intermediate frequency range ($\omega=$1-10 THz) for $\Lambda_{m}=50a$,
the attenuation length is almost a constant with $\lambda\sim3$ $\mu$m
or about two orders of magnitude larger than $\Lambda_{m}$.}

\label{Fig:CorrelatedMassTransmittance}
\end{figure}

\begin{figure}
\includegraphics[width=8.5cm]{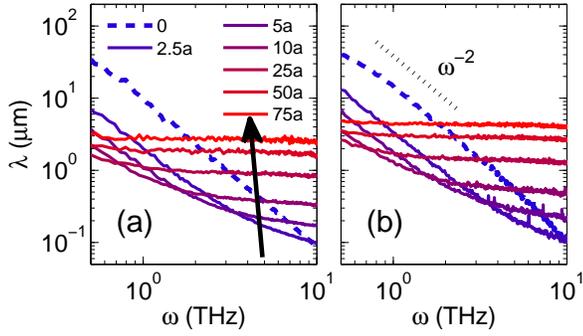}

\caption{(Color online) Plot of the attenuation length $\lambda$ computed
from (a) the transfer matrix method and (b) the transmittance function,
as a function of frequency $\omega$ for different values of the mass
correlation length ($\Lambda_{m}/a=0$, 2.5, 5, 10, 25, 50 and 75)
in the frequency range of 0.5 to 10 THz. The attenuation length in
(a) is slightly smaller than in (b). However, the change in $\lambda$
with $\Lambda_{m}$ is qualitatively the same for both methods. The
arrow indicates the direction of increasing $\Lambda_{m}$. In the
uncorrelated case ($\Lambda_{m}=0$), the attenuation length scales
as $\lambda\propto\omega^{-2}$. Mass correlation increases (decreases)
the attenuation length at high (low) frequencies. In the 0.5 to 10
THz range, the attenuation length for $\Lambda_{m}=75a$ (or 41.25
nm) is around 3 $\mu$m, about 2 orders of magnitude larger. We interpret
this as the formation of semi-extended vibrational states.}

\label{Fig:AttenuationLengths}
\end{figure}

\section{Results}

\subsection{Length dependence of phonon transmittance\emph{ }}

Figure\ \ref{Fig:CorrelatedMassTransmittance} shows the transmittance
as a function of $L$ and $\omega$ for (a) $\Lambda_{m}=0$, (b)
$\Lambda_{m}=10a$, and (c) $\Lambda_{m}=50a$. In the uncorrelated
case in Fig.\ \ref{Fig:CorrelatedMassTransmittance}(a), the transmittance
function scales as $\langle\Xi(\omega,L)\rangle=\exp\left[-L/\lambda(\omega)\right]$,
where $\lambda\propto\omega^{-2}$. In Fig.\ \ref{Fig:CorrelatedMassTransmittance}(b),
when $\Lambda_{m}=10a$, the attenuation length still scales as $\lambda\propto\omega^{-2}$
at low frequencies but for $\omega>1$ THz, it deviates from that
behavior as it varies weakly with $\omega$ at a characteristic length
scale of $\lambda\sim$ 0.5 $\mu$m. Figure\ \ref{Fig:CorrelatedMassTransmittance}(c)
also shows a frequency regime in which $\lambda$ is weakly $\omega$
dependent in the range 0.5 to 10 THz. The characteristic attenuation
length scale is, however, at $\sim2.5$ $\mu$m. The plots in Fig.\ \ref{Fig:CorrelatedMassTransmittance}
show that a finite $\Lambda_{m}$ alters the attenuation length at
high frequencies. To see this more clearly, we plot $\lambda(\omega)$
in Fig.\ \ref{Fig:AttenuationLengths} estimated numerically from
(a) the Lyapunov exponent {[}$\lambda=a(2\gamma)^{-1}${]} and (b)
the attenuation of the transmittance function {[}$\lambda(\omega)=\int dL\ \langle\Xi(\omega,L)\rangle${]}.
There is good agreement between the $\lambda(\omega)$ values computed
from the two methods with the $\lambda$ from (b) being slightly larger. 

Two trends are immediately clear in Fig.\ \ref{Fig:AttenuationLengths}.
Firstly, a nonzero $\Lambda_{m}$ causes the attenuation length to
deviate from $\lambda_{0}$ ($\Lambda_{m}=0$). At higher frequencies,
transmittance is enhanced, i.e.\emph{,} $\lambda>\lambda_{0}$. As
$\Lambda_{m}$ increases, the frequency range in which $\lambda$
is weakly $\omega$-dependent becomes wider. Furthermore, $\lambda\gg\Lambda_{m}$,
and $\lambda(\omega)$ also increases with $\Lambda_{m}$. Secondly,
transmittance is \emph{reduced} at low frequencies, i.e., $\lambda<\lambda_{0}$.
The low-frequency phononic opacity is increased by the mass correlation.
The weak $\omega$ dependence of the attenuation length of the higher-frequency
modes also implies that these modes are `semi-extended' and participate
in ballistic heat conduction when the lattice size is comparable to
their attenuation lengths. Physically, the short-range order in the
mass distribution results in the partial \emph{delocalization} of
the higher-frequency modes but greater \emph{localization} of the
low-frequency modes. This alters the relative participation of the
phonon modes in length-dependent heat conduction.

\begin{figure}
\includegraphics[width=8.5cm]{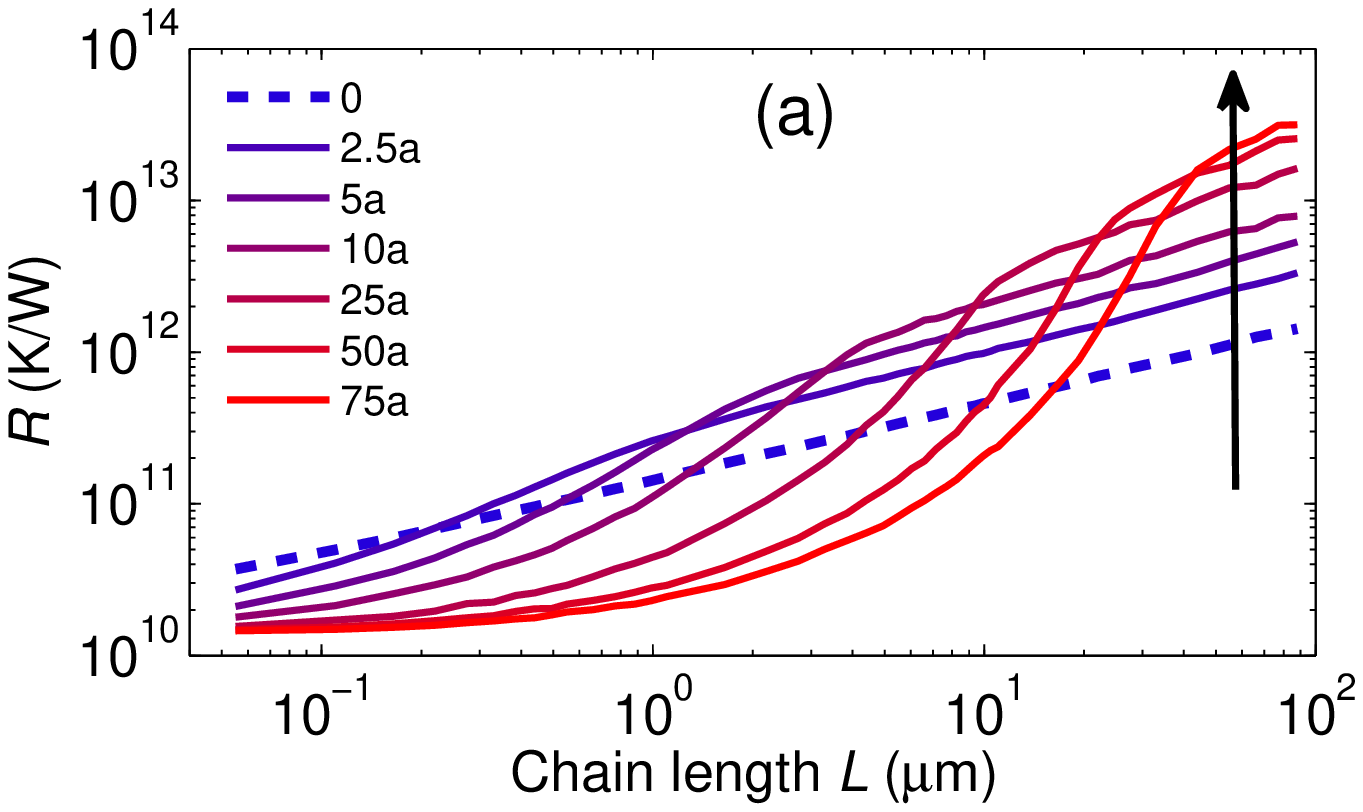}

\includegraphics[width=8.5cm]{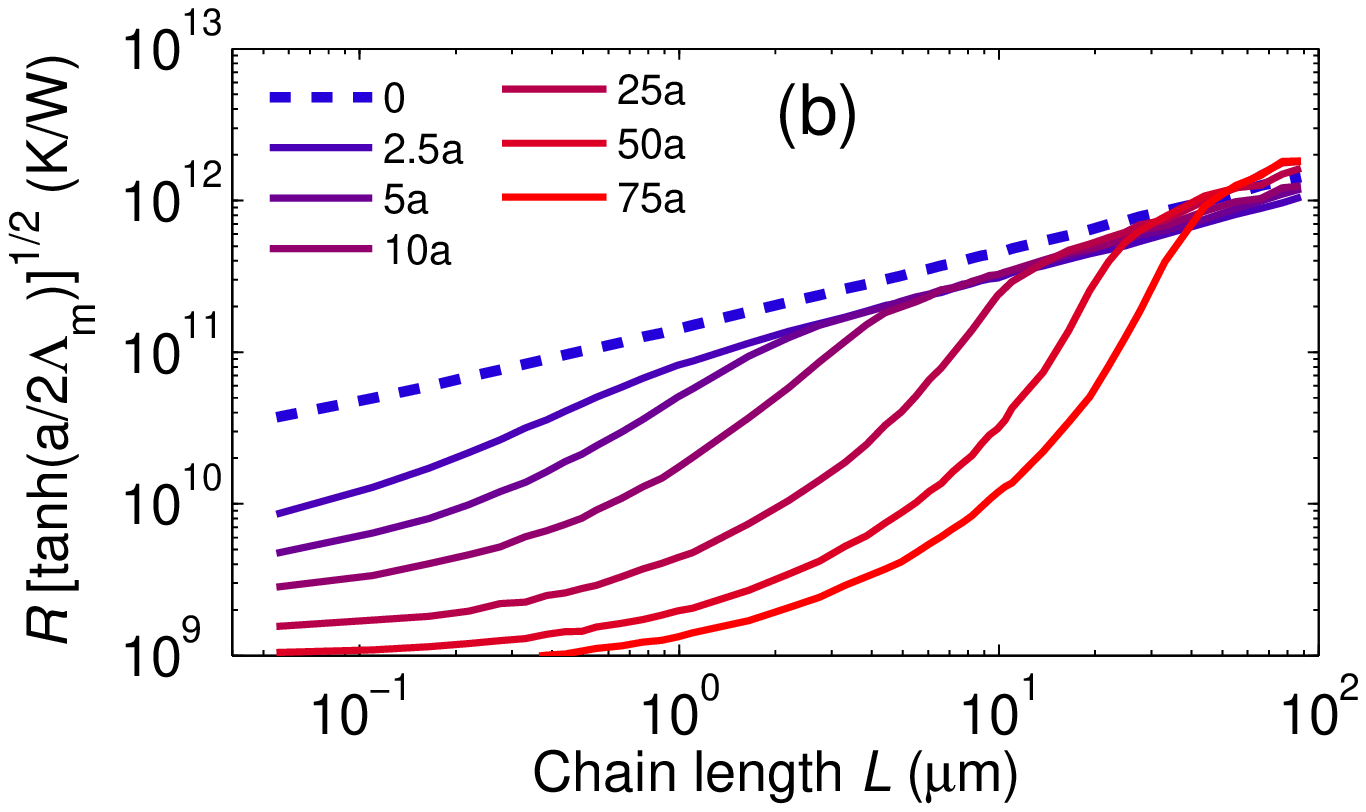}

\caption{(Color online) (a) Plot of the thermal resistance $R$ as a function
of chain length $L$ for for different values of the mass correlation
length ($\Lambda_{m}/a=0$, 2.5, 5, 10, 25, 50 and 75) in the $T\rightarrow\infty$
limit. The arrow indicates the direction of increasing $\Lambda_{m}$.
For $\Lambda_{m}=0$, i.e.\emph{,} the uncorrelated case, $R$ scales
as $R\propto L^{0.5}$, as expected. However, when mass correlation
is finite, the thermal resistance relative to the uncorrelated case
decreases at small $L$. As $L\rightarrow0$, the resistance converges
to a constant, implying the existence of ballistic heat conduction.
At large $L$, the thermal resistance goes up as we increase the mass
correlation length. (b) Plot of the rescaled thermal resistance normalized
by $[\tanh(\frac{a}{2\Lambda_{m}})]^{1/2}$ as a function of $L$.
In the $L\rightarrow\infty$ limit, $R[\tanh(\frac{a}{2\Lambda_{m}})]^{1/2}$
for different values of $\Lambda_{m}$ appears to converge. }

\label{Fig:ThermalResistanceVsLength}
\end{figure}

\subsection{Thermal resistance}

To determine the effect of the altered transmittance on heat conduction,
we compute the length-dependent thermal resistance $R(L)$ using the
Landauer formula:\ \cite{WZhang:NHT07,NMingo:Springer09}
\begin{equation}
R(L)=\left[\frac{1}{2\pi}\int_{0}^{\omega_{L}}\hbar\omega\frac{df(\omega)}{dT}\langle\Xi(\omega,L)\rangle\ d\omega\right]^{-1}\label{Eq:LandauerFormula}
\end{equation}
where $T$ and $\hbar$ are respectively the temperature and Planck
constant, and $f(\omega)=[\exp(\frac{\hbar\omega}{k_{B}T})-1]^{-1}$is
the Bose-Einstein occupation factor. $\omega_{L}$ is the cutoff frequency
determined by the pristine lead ($\omega_{L}=\sqrt{4k/m_{B}}$). $\langle\Xi(\omega,L)\rangle$
is derived from our nonequilibrium Green's function (NEGF) computation.
In the high-temperature limit, Eq.\ (\ref{Eq:LandauerFormula}) becomes
$\lim_{T\rightarrow\infty}R(L)=\left[\frac{k_{B}}{2\pi}\int_{0}^{\infty}\langle\Xi(\omega,L)\rangle\ d\omega\right]^{-1}$. 

The high-temperature thermal resistance is shown in Fig.\ \ref{Fig:ThermalResistanceVsLength}(a)
for different values of $\Lambda_{m}$ ($\Lambda_{m}/a=0$, 2.5, 5,
10, 25, 50 and 75). We take the $\Lambda_{m}=0$ curve (which we term
$R_{0}$) to be the baseline for comparison. $R_{0}(L)$ scales as
$L^{0.5}$ at all values of $L$, as predicted for uncorrelated disorder.
However, for correlated disorder ($\Lambda_{m}\neq0$), $R(L)$ deviates
from the $R\propto L^{0.5}$ behavior. At small $L$, $R<R_{0}$ and
is weakly dependent on $L$, implying that the system conducts heat
ballistically. This is due to the enhanced attenuation length and
transmittance of the high-frequency modes relative to the uncorrelated
case. We also observe that $\lim_{L\rightarrow0}R$ converges to the
same value ($R=\frac{\sqrt{8}\pi}{k_{B}\omega_{L}}\approx1.2\times10^{10}$K/W)
regardless of the correlation length. This value is determined by
the cutoff frequency and impurity mass. As $L\rightarrow\infty$,
the relative contribution of the low-frequency modes grows. Thus,
$R$ increases rapidly and exceeds $R_{0}$, scaling as $L^{0.5}$,
because of greater low-frequency phononic opacity. 

We quantify the dependence of the thermal resistance on mass correlation
here in the $L\rightarrow\infty$ limit. As noted earlier, the ratio
$\lim_{L\rightarrow\infty}R/R_{0}$ grows monotonically with $\Lambda_{m}$.
The thermal resistance in the $\Lambda_{m}=0$ case,\ \cite{ZYOng:Manuscript14_MassDisorder}
\begin{equation}
\lim_{L\rightarrow\infty}R_{0}(L)=\sqrt{\frac{4\pi\langle\delta m^{2}\rangle L}{ka\langle m\rangle k_{B}^{2}}}\ ,\label{Eq:DisorderedThermalResistance}
\end{equation}
is proportional to the local mass fluctuation $\sqrt{\langle\delta m^{2}\rangle}$.
Equation\ (\ref{Eq:DisorderedThermalResistance}) is generalized
(see Appendix\ \ref{Appendix:ThermalResistance} for the derivation)
by replacing $\langle\delta m^{2}\rangle$ with $\sum_{x}\langle\delta m(x)\delta m(0)\rangle$,
which sums over the mass fluctuation across atomic sites, to yield
\begin{equation}
\lim_{L\rightarrow\infty}R(\Lambda_{m},L)=\sqrt{\frac{4\pi\sum_{x}\langle\delta m(x)\delta m(0)\rangle L}{ka\langle m\rangle k_{B}^{2}}}\label{Eq:CorrelatedThermalResistance}
\end{equation}
where $\langle\delta m(x)\delta m(0)\rangle=\langle\delta m^{2}\rangle e^{-|x|/\Lambda_{m}}$,
giving us $\lim_{L\rightarrow\infty}R(\Lambda_{m},L)=R_{0}(L)[\tanh(\frac{a}{2\Lambda_{m}})]^{-1/2}$
since $\lim_{N\rightarrow\infty}\sum_{x=-Na}^{Na}e^{-|x|/\Lambda_{m}}=1/\tanh(\frac{a}{2\Lambda_{m}})$.
We recover Eq.\ (\ref{Eq:DisorderedThermalResistance}) from Eq.\ (\ref{Eq:CorrelatedThermalResistance})
for $\Lambda_{m}=0$, while for large $\Lambda_{m}$, we obtain $\lim_{L\rightarrow\infty}R(\Lambda_{m},L)\propto\sqrt{\Lambda_{m}}$.
It should be noted that the results in Eqs.\ (\ref{Eq:DisorderedThermalResistance})
and (\ref{Eq:CorrelatedThermalResistance}) are derived assuming that
the semi-infinite homogeneous leads act as heat baths and are coupled
to the lattice via the same harmonic spring terms as those between
the atoms in the disordered lattice. We plot the normalized thermal
resistance $R[\tanh(\frac{a}{2\Lambda_{m}})]^{1/2}$ in Fig.\ \ref{Fig:ThermalResistanceVsLength}(b).
In the $L\rightarrow\infty$ limit, $R[\tanh(\frac{a}{2\Lambda_{m}})]^{1/2}$
for different values of $\Lambda_{m}$ converges to $R_{0}(L)$, suggesting
that Eq.\ (\ref{Eq:CorrelatedThermalResistance}) captures the effect
of mass correlation on thermal resistance. 

\begin{figure}
\includegraphics[width=8.5cm]{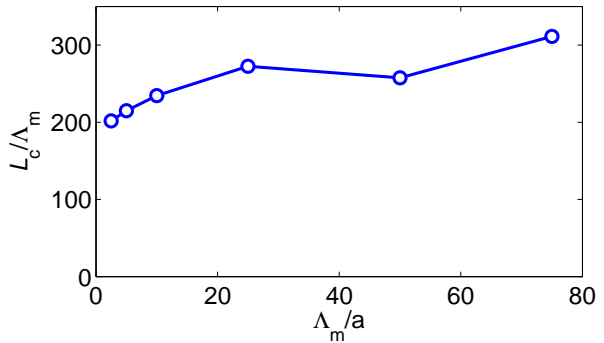}

\caption{(Color online) Plot of the ratio of the crossover length $L_{c}$
to the mass correlation length $\Lambda_{m}$ as a function of $\Lambda_{m}$.
The crossover length is the minimum attenuation length at which $\lambda(\Lambda_{m},\omega)>\tanh(\frac{a}{2\Lambda_{m}})\lambda(\Lambda_{m}=0,\omega)$,
i.e.\emph{,} $\lambda$ deviates from the $\lambda\propto\omega^{-2}$
behavior. We observe that $L_{c}$ is about 250 times (or two orders
of magnitude) larger than $\Lambda_{m}$.}

\label{Fig:CorrelationVsCrossover}
\end{figure}

\subsection{Crossover length\emph{ }}

At low frequency, the attenuation length for the uncorrelated case
is $\lim_{\omega\rightarrow0^{+}}\lambda_{0}(\omega)=4ka\langle m\rangle/(\langle\delta m^{2}\rangle\omega^{2})$.\ \cite{ZYOng:Manuscript14_MassDisorder}
The corresponding attenuation length for correlated disorder is $\lim_{\omega\rightarrow0^{+}}\lambda(\Lambda_{m},\omega)=\tanh(\frac{a}{2\Lambda_{m}})\lambda(\Lambda_{m}=0,\omega)$
(see Appendix\ \ref{Appendix:AnalyticalEstimate} for the derivation).
In the $\Lambda_{m}\rightarrow0$ limit, we recover $\lim_{\omega\rightarrow0^{+},\Lambda_{m}\rightarrow0}\lambda(\Lambda_{m},\omega)=\lambda_{0}(\omega)$
while for large $\Lambda_{m}$, we obtain $\lim_{\omega\rightarrow0^{+}}\lambda(\Lambda_{m},\omega)\propto1/\Lambda_{m}$,\emph{
}i.e., the attenuation length is inversely proportional to the mass
correlation length.

We define the crossover length scale $L_{c}$ as the minimum attenuation
length at which $\lambda(\Lambda_{m},\omega)\geq\tanh(\frac{a}{2\Lambda_{m}})\lambda(\Lambda_{m}=0,\omega)$,
i.e., the smallest attenuation length for the correlated case is greater
than or equal to the rescaled attenuation length for the uncorrelated
length. Physically, $L_{c}$ sets the length scale above which Eq.\ (\ref{Eq:CorrelatedThermalResistance})
describes the thermal resistance and the $R\propto L^{-0.5}$ scaling
behavior applies as the contribution of the semi-extended high-frequency
modes vanishes. For $L\ll L_{c}$, the system becomes thermally transparent
{[}i.e., $\langle\Xi(\omega,L)\rangle\approx1${]} and heat conduction
becomes quasiballistic given the participation of the semi-extended
states. Figure\ \ref{Fig:CorrelationVsCrossover} shows the numerically
computed ratio $L_{c}/\Lambda_{m}$ as a function of $\Lambda_{m}$.
We observe that $L_{c}$ averages about 250 times (or two orders of
magnitude) larger than $\Lambda_{m}$ over the range of correlation
lengths considered. This demonstrates that ballistic heat conduction
over a given length may be associated with a mass correlation that
is two orders of magnitude smaller.

\section{Discussion }

Our simulation results imply that heat conduction in alloys is not
merely a function of the mass difference and the relative concentrations\ \cite{ZYOng:Manuscript14_MassDisorder}
but also depends on the mass distribution. In the perfect alloy, the
solubility of the component species is assumed to be equal and the
atoms are randomly distributed across lattice sites. However, differences
in miscibility\  and growth kinetics\ \cite{QYu:PRB96_Solidification}
may lead to phase segregation or the formation of domains,\ \cite{JNAqua:PhysReports13_Growth}
introducing short-range order in the mass distribution and forming
semi-extended high-frequency modes that are spatially larger than
the correlation length. Hence, it is possible for an alloy 1D nanostructure
to conduct heat ballistically at a finite length scale that is much
larger than the average domain size. The key point here is that the
short-range order drastically increases the localization length of
these high-frequency modes, allowing them to contribute substantially
to heat conduction. Furthermore, the weakly frequency-dependent localization
length of the high-frequency modes sets an \emph{intrinsic} length
scale below which a change in the length of the lattice does not affect
the phonon contribution to the heat current, i.e., the coherent heat
conduction can be ballistic over a finite length. 

The interplay between ordering and heat conduction has been observed
in planar SiGe superlattices.\ \cite{PChen:PRL14} Chen and co-workers
found that the thermal conductivity is considerably lower in SiGe
with pure layers separated by sharp interfaces than it is in homogeneous
SiGe alloys. This is attributed to the higher scattering efficiency
of low-frequency phonons by the pure domains created from the long-range
compositional order. Moreover, the thermal conductivity is further
reduced when there is some grading in the concentration profile at
the interface which introduces short-range disorder and enhances the
scattering of higher-frequency phonons. This variation in the thermal
conductivity and frequency-dependent scattering efficiency in the
superlattice structure highlights the role of correlated disorder
in heat conduction.

\begin{figure}
\includegraphics[width=8.5cm]{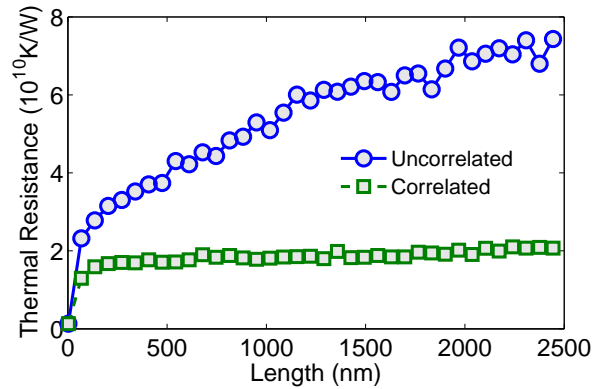}

\caption{(Color online) Thermal resistance as a function of length for a strongly
mass-disordered 1.1 nm-wide rectangular Si nanowire with uncorrelated
(circle) and correlated (square) disorder. Fifty percent of the atoms
have their masses set equal to that of Ge while the rest have theirs
set to that of Si. The correlation length in the correlated disorder
case is 11 nm.}
\label{Fig:SiGeNanowire}
\end{figure}

To elucidate the effect of correlated disorder in quasi-one-dimensional
nanostructures, we calculate the thermal resistance as a function
of length for a model 1.1 nm-wide rectangular Si nanowire that has
a 50:50 mix of isotopes at 300 K. To mimic the effect of mass disorder
in a Si$_{0.5}$Ge$_{0.5}$ nanowire, we set the atomic mass of the
lighter and heavier isotope to be equal to that of Si and Ge, respectively.
The transmittance is calculated using the NEGF method and then used
to compute the thermal resistance {[}see Eq.\ (\ref{Eq:LandauerFormula}){]}
as a function of temperature. Figure\ \ref{Fig:SiGeNanowire} shows
the results averaged over ten instances of disorder for uncorrelated
and correlated ($\Lambda_{m}=20a$ or 11 nm) disorder for nanowires
up to 2.5 $\mu$m. We observe that the thermal resistance in the uncorrelated
case is significantly more length dependent than in the correlated
case where the weak length dependence suggests ballistic heat conduction
and is reminiscent of the experimental results observed in Ref.\ \cite{TKHsiao:NatNano13_Observation}.
Qualitatively, the length dependence is also similar to that in Fig.\ \ref{Fig:ThermalResistanceVsLength}(a)
where a finite correlation length also leads to micrometer-scale quasiballistic
heat conduction. While our model represents a very drastic idealization
of real SiGe nanowires with anharmonic phonon-phonon interactions
and boundary scattering completely neglected, it does suggest that
even some short-range order can have a dramatic effect on heat conduction
in quasi-one-dimensional nanostructures, especially when phonon transport
is limited by alloy scattering.

Our results also highlight the role of short-range order in the alloy
structure on phonon-mediated heat conduction, a factor which has not
been given significant consideration in the literature. It is known
that the attenuation length $\lambda$ in the transmittance is proportional
to the mean free path via the Thouless relation.\ \cite{PWAnderson:PRB80_NewMethod,ISavic:PRL08_CNTLocalization}
Thus, if the positions of the alloy atoms are correlated, our results
imply that the mean free paths of the low-frequency phonons are accordingly
reduced while those of the high-frequency modes are extended, modifying
the phonon contribution balance to the thermal conductivity. One potential
application of this is the manipulation of the thermal conductivity
of alloys by controlling the spatial distribution of the alloy components.
One approach to the control of heat flow has been to engineer the
spectrum of the phonons in order to select for phonons of specific
frequencies that can be managed by a periodic structure.\ \cite{MMaldovan:PRL13_Narrow}
In particular, the fabrication of thermocrystals\ \cite{MMaldovan:PRL13_Narrow}
requires the suppression of the very-low-frequency phonons which may
be realized by using a SiGe alloy with finite short-range order.

\section{Conclusion\emph{ }}

We have simulated heat conduction in the 1D alloy-like RBL for correlated
and uncorrelated mass disorder. We find that correlated mass disorder
enhances (reduces) the high-frequency (low-frequency) phonon transmittance,
leading to ballistic heat conduction that persists over a length scale
that is two orders of magnitude larger than the mass correlation length.
However, in very large systems, correlated mass disorder increases
the thermal resistance of the system because of the greater low-frequency
phononic opacity.

\section*{Acknowledgment}

The authors gratefully acknowledge the financial support from the
Agency for Science, Technology and Research (A{*}STAR), Singapore
and the use of computing resources at the A{*}STAR Computational Resource
Centre, Singapore.

\appendix

\section{Analytical estimate of low-frequency inverse localization length
for correlated mass disorder\label{Appendix:AnalyticalEstimate}}

We derive an analytical expression for the inverse localization length
in the $\omega\rightarrow0^{+}$ limit for a given mass correlation,
using the transfer matrix method.\ \cite{BDerrida:JDePhysique87,GJKissel:PRA91_Localization}
The transmittance through a lattice of $N$ atoms can be written as
the absolute square of the transmission function, i.e., $\Xi(\omega)=|\tau_{N}|^{2}$,
where $\tau_{N}$ is the transmission function with the amplitude
$\lim_{N\rightarrow\infty}|\tau_{N}|=\exp[-\gamma(\omega)N]$. The
inverse localization length $\gamma$ (also commonly known as the
Lyapunov exponent) at frequency $\omega$ is thus defined as 
\begin{equation}
\gamma(\omega)=\lim_{N\rightarrow\infty}\frac{1}{N}\ln|\tau_{N}|\ ,\label{Eq:LyapunovExponent}
\end{equation}
and $|\tau_{N}|$ can be computed from the norm of the product of
$N$ transfer matrices\ \cite{ADhar:PRL01_DisorderedHarmonicChain},
i.e., 
\begin{equation}
|\tau_{N}|=\left\Vert \prod_{p=1}^{N}T_{p}\right\Vert \ .\label{Eq:TransferMatricesProduct}
\end{equation}
The transmittance attenuation length is $\lambda(\omega)=a[2\gamma(\omega)]^{-1}$,
where $a$ is the interatomic spacing. $\lambda$ can be directly
computed numerically from Eqs.\ (\ref{Eq:LyapunovExponent}) and
(\ref{Eq:TransferMatricesProduct}), as in Refs.\ \cite{FABFDeMoura:PRB03_DelocalizationCorrelatedRandomMasses,FABFDeMoura:PRB06_Aperiodic},
or estimated analytically in the $\omega\rightarrow0$ limit as follows.

The $p$-th transfer matrix in Eq.\ (\ref{Eq:TransferMatricesProduct})
is defined as
\begin{equation}
T_{p}=\left(\begin{array}{cc}
2-x_{p} & -1\\
1 & 0
\end{array}\right)\ ,\label{Eq:TransferMatrix}
\end{equation}
where $x_{p}$ is the dimensionless random variable corresponding
to the $p$-th atom in the chain such that $x_{p}=m_{p}\omega^{2}/k$
and $0\leq x_{p}\leq2$, and is a particular formulation of the equation
of motion for the spatial displacement of the $p$-th atom.\ \cite{FABFDeMoura:PRB03_DelocalizationCorrelatedRandomMasses}

For weak disorder or at low frequencies, we can find an analytical
approximation for the localization length following the method described
in Refs.\ \cite{BDerrida:JDePhysique87,GJKissel:PRA91_Localization}.
Let $x_{p}=\langle x\rangle+\delta x_{p}$ where $\langle x_{p}\rangle=\langle x\rangle$
and $\langle\delta x_{p}\rangle=0$. The expression in Eq.\ (\ref{Eq:TransferMatrix})
can thus be linearized as
\[
T_{p}=\left(\begin{array}{cc}
2-\langle x\rangle & -1\\
1 & 0
\end{array}\right)+\left(\begin{array}{cc}
-1 & 0\\
0 & 0
\end{array}\right)\delta x_{p}\ .
\]
We choose an eigenvector transformation $U$, such that $\tilde{T}_{p}=U^{\dagger}T_{p}U$,
to obtain
\[
\tilde{T}_{p}=\left(\begin{array}{cc}
e^{i\theta} & 0\\
0 & e^{-i\theta}
\end{array}\right)+\frac{1}{2i\sin\theta}\left(\begin{array}{cc}
-e^{i\theta} & -e^{-i\theta}\\
e^{i\theta} & e^{-i\theta}
\end{array}\right)\delta x_{p}
\]
where $\theta$ is defined by $\cos\theta=1-\frac{1}{2}\langle x\rangle$
and $\sin\theta=[\langle x\rangle-\langle x\rangle^{2}/4]^{1/2}$.
For notational convenience, we define 
\[
A=\left(\begin{array}{cc}
e^{i\theta} & 0\\
0 & e^{-i\theta}
\end{array}\right)
\]
\[
B=\frac{1}{2i\sin\theta}\left(\begin{array}{cc}
-e^{i\theta} & -e^{-i\theta}\\
e^{i\theta} & e^{-i\theta}
\end{array}\right)
\]
so that we can write the $p$-th transfer matrix as $\tilde{T}_{p}=A+B\delta x_{p}$.
The $N$-product of the transfer matrices is 
\begin{equation}
\prod_{p=1}^{N}\tilde{T}_{p}=(A+B\delta x_{1})\ldots(A+B\delta x_{N})\ .\label{Eq:TransferMatrixProduct}
\end{equation}

The expression in Eq.\ (\ref{Eq:TransferMatrixProduct}) can be systematically
expanded in powers of $\delta x_{p}$. The zeroth-, first- and second-order
terms are
\begin{equation}
A^{N}\ ,
\end{equation}
\begin{equation}
\sum_{p=1}^{N}A^{p-1}BA^{N-p}\delta x_{p}
\end{equation}
and

\begin{equation}
\sum_{p=1}^{N-1}\sum_{q=p+1}^{N}A^{p-1}BA^{q-p-1}BA^{N-q}\delta x_{p}\delta x_{q}\ ,
\end{equation}
 respectively. To find the low-frequency scaling behavior, we need
only consider the zeroth- and first-order terms. Thus, we write Eq.\ (\ref{Eq:TransferMatrixProduct})
as
\begin{equation}
\prod_{p=1}^{N}\tilde{T}_{p}\approx A^{N}+\sum_{p=1}^{N}A^{p-1}BA^{N-p}\delta x_{p}\ .\label{Eq:TruncatedProduct}
\end{equation}
The (1,1) term of the $2\times2$ matrix in Eq.\ (\ref{Eq:TruncatedProduct})
is equal to $1/\tau_{N}$, and after some algebra, can be written
as $e^{iN\theta}[1-(2i\sin\theta)^{-1}\sum_{p=1}^{N}\delta x_{p}]$.
Hence, 
\[
|\tau_{N}|^{2}=\left[1+i\frac{\cos N\theta}{\sin\theta}\sum_{p=1}^{N}\delta x_{p}+\frac{1}{4\sin^{2}\theta}\left(\sum_{p=1}^{N}\delta x_{p}\right)^{2}\right]^{-1}
\]
and its logarithm can be approximated as 
\begin{equation}
\ln|\tau_{N}|^{2}\approx-\frac{i\cos N\theta}{\sin\theta}\sum_{p=1}^{N}\delta x_{p}-\frac{1}{4\sin^{2}\theta}\left(\sum_{p=1}^{N}\delta x_{p}\right)^{2}\ .\label{Eq:LogTransmissionSquared}
\end{equation}
The localization length is defined as $\gamma=-\lim_{N\rightarrow\infty}\frac{1}{N}\langle\ln|\tau_{N}|\rangle$.
The ensemble averaging $\langle\ldots\rangle$ removes the terms linear
in $\delta x_{n}$ in Eq.\ (\ref{Eq:LogTransmissionSquared}). Bearing
in mind that we are only considering the low-frequency scaling behavior,
the expression for the inverse localization length is 
\begin{eqnarray}
\lim_{\omega\rightarrow0^{+}}\gamma(\omega) & = & \lim_{N\rightarrow\infty}\frac{1}{8N\sin^{2}\theta}\left\langle \left(\sum_{p=1}^{N}\delta x_{p}\right)^{2}\right\rangle \nonumber \\
 & = & \lim_{N\rightarrow\infty}\frac{1}{8N\sin^{2}\theta}\sum_{p=1}^{N}\sum_{q=1}^{N}\langle\delta x_{p}\delta x_{q}\rangle\ .\label{Eq:InverseLocalizationLength}
\end{eqnarray}
In the correlated case, 
\[
\langle\delta x_{p}\delta x_{q}\rangle=\langle\delta x^{2}\rangle\exp(-|x_{p}-x_{q}|/\Lambda_{m})
\]
and the double sum in Eq.\ (\ref{Eq:InverseLocalizationLength})
becomes 
\[
\sum_{p=1}^{N}\sum_{q=1}^{N}\langle\delta x_{p}\delta x_{q}\rangle\approx N\langle\delta x^{2}\rangle\coth\left(\frac{a}{2\Lambda_{m}}\right)\ .
\]
Therefore, the inverse localization length is
\begin{equation}
\lim_{\omega\rightarrow0^{+}}\gamma(\omega)\approx\frac{\langle\delta m^{2}\rangle\omega^{2}}{8k\langle m\rangle}\coth\left(\frac{a}{2\Lambda_{m}}\right)\ .\label{Eq:InverseLocalizationLength-2}
\end{equation}
Note that in the uncorrelated case, the correlation length is $\Lambda_{m}=0$
and the expression in Eq.\ (\ref{Eq:InverseLocalizationLength-2})
yields
\[
\lim_{\omega\rightarrow0^{+}}\gamma(\omega)=\frac{\langle\delta m^{2}\rangle\omega^{2}}{8k\langle m\rangle}
\]
as expected. We remind the reader that the analytical expression in
Eq.\ (\ref{Eq:InverseLocalizationLength-2}) only applies at low
frequencies. At higher frequencies, this approximation fails and the
Lyapunov exponent is calculated directly from Eq.\ (\ref{Eq:LyapunovExponent}).

\section{Thermal resistance in the thermodynamic ($L\rightarrow\infty$) limit
for correlated mass disorder\label{Appendix:ThermalResistance}}

Given the expression for $\gamma(\omega)$ in Eq.\ (\ref{Eq:InverseLocalizationLength-2}),
we can obtain the expression for the thermal resistance in the $L\rightarrow\infty$
limit. The phonon transmittance is 
\[
\lim_{L\rightarrow\infty}\langle\Xi(\omega,L)\rangle=\exp[-2\gamma(\omega)L/a]
\]
and the high-temperature length-dependent thermal conductance is
\begin{equation}
\lim_{L\rightarrow\infty}\sigma(L)=\frac{k_{B}}{2\pi}\int_{0}^{\infty}\langle\Xi(\omega,L)\rangle\ d\omega\ .\label{Eq:ThermodynamicLimitConductance}
\end{equation}
The integrand in Eq.\ (\ref{Eq:ThermodynamicLimitConductance}) vanishes
as $\omega\rightarrow\infty$ and only its low-frequency part contributes
to the integral, allowing us to use Eq.\ (\ref{Eq:InverseLocalizationLength-2}).
Thus, the explicit expression for Eq.\ (\ref{Eq:ThermodynamicLimitConductance})
is 
\[
\lim_{L\rightarrow\infty}\sigma(L)=\left[\frac{ka\langle m\rangle k_{B}^{2}}{4\pi\langle\delta m\rangle^{2}L}\tanh\left(\frac{a}{2\Lambda_{m}}\right)\right]^{1/2}\ .
\]
As a function of the mass correlation length, the asymptotic ($L\rightarrow\infty$)
expression for the thermal resistance $R=1/\sigma$ is
\[
\lim_{L\rightarrow\infty}R(\Lambda_{m},L)=\frac{R_{0}(L)}{\sqrt{\tanh\left(\frac{a}{2\Lambda_{m}}\right)}}
\]
where $R_{0}(L)=\lim_{L\rightarrow\infty}R(\Lambda_{m}=0,L)\ .$ 

\bibliographystyle{apsrev4-1}
\bibliography{PhononNotes}

\end{document}